Research Article

arXiv

# Sensory Robustness through Top-Down Feedback and Neural Stochasticity in Recurrent Vision Models

Antonino Greco[1,2,3,4,✉], Marco D'Alessandro[5], Karl J. Friston[6], Giovanni Pezzulo[5] & Markus Siegel[1,2,3,4,7,✉]

[1] Department of Neural Dynamics and Magnetoencephalography, Hertie Institute for Clinical Brain Research, University of Tübingen, Tübingen, Germany

[2] Werner Reichardt Centre for Integrative Neuroscience, University of Tübingen, Tübingen, Germany

[3] MEG Center, University of Tübingen, Tübingen, Germany

[4] Center for Bionic Intelligence Tübingen Stuttgart (BITS), Tübingen, Germany

[5] Institute of Cognitive Sciences and Technologies, National Research Council, Rome, Italy

[6] Department of Imaging Neuroscience, Queen Sq., Institute of Neurology, University College London, London, United Kingdom

[7] German Center for Mental Health (DZPG), Tübingen, Germany

✉ Corresponding authors: Antonino Greco (antonino.greco@uni-tuebingen.de) & Markus Siegel (markus.siegel@uni-tuebingen.de)

## Abstract

**Biological systems leverage top-down feedback for visual processing, yet most artificial vision models succeed in image classification using purely feedforward or recurrent architectures, calling into question the functional significance of descending cortical pathways. Here, we trained convolutional recurrent neural networks (ConvRNN) on image classification in the presence or absence of top-down feedback projections to elucidate the specific computational contributions of those feedback pathways. We found that ConvRNNs with top-down feedback exhibited remarkable speed-accuracy trade-off and robustness to noise perturbations and adversarial attacks, but only when they were trained with stochastic neural variability, simulated by randomly silencing single units via dropout. By performing detailed analyses to identify the reasons for such benefits, we observed that feedback information substantially shaped the representational geometry of the post-integration layer, combining the bottom-up and top-down streams, and this effect was amplified by dropout. Moreover, feedback signals coupled with dropout optimally constrained network activity onto a low-dimensional manifold and encoded object information more efficiently in out-of-distribution regimes, with top-down information stabilizing the representational dynamics at the population level. Together, these findings uncover a dual mechanism for resilient sensory coding. On the one hand, neural stochasticity prevents unit-level co-adaptation albeit at the cost of more chaotic dynamics. On the other hand, top-down feedback harnesses high-level information to stabilize network activity on compact low-dimensional manifolds.**

## Acknowledgements

This study was supported by the European Research Council (ERC; https://erc.europa.eu/, CoG 864491 to M.S.; CoG 820213 to G.P.), by the German Research Foundation (DFG; https://www.dfg.de/, projects 276693517 (SFB 1233; to M.S.) and SI 1332/6-1 (SPP 2041; to M.S.)) and by the Wellcome Trust (https://wellcome.org/; Ref: 226793/Z/22/Z; to K.F.).

## Introduction

Biological visual systems leverage top-down feedback for visual processing[1–4]. In the brain, feedback projections from higher to lower visual areas abound, influencing processing from contour integration to attentional selection and perceptual inference[5–12]. Computational hypotheses have posited that feedback serves to refine noisy sensory inputs, transmits predictive signals for hierarchical inference and resolves ambiguity through contextual priors[13–24].

In contrast, deep learning has revolutionized computer vision with feedforward convolutional neural networks (CNN) reaching human performance on large-scale image-classification benchmarks[25–28]. Furthermore, feedforward CNNs predict neural responses in the primate ventral stream, accounting for both single-unit and population-level activity patterns[29–40]. However, substantial evidence indicates that recurrence is essential for aligning models with ventral-stream object-recognition behavior[41,42]. Recurrent vision architectures are typically implemented as convolutional recurrent neural networks (ConvRNNs) and incorporate two forms of projections: lateral (recurrent) connections, which reinject each layer's previous hidden state back into itself, and top-down feedback projections, which project information from higher to lower layers across time.

A growing body of work has demonstrated that ConvRNNs outperform purely feedforward CNNs on out-of-distribution challenges such as occlusion, noise perturbation, and adversarial attacks[43–52]. However, these studies most often compare feedforward models against architectures that combine both lateral and feedback connections, leaving the individual roles of each pathway unclear. Notable exceptions include Kietzmann et al.[42], who systematically ablated recurrent and feedback connections and demonstrated that, unexpectedly, lateral connections contribute more than top-down feedback to both object-recognition performance and alignment with neural data. Moreover, Spoerer et al.[44] directly contrasted lateral recurrence and top-down feedback and reported mixed outcomes across occlusion and noise perturbations, though their models differed in parameter count and were primarily designed to compare feedforward versus recurrent architectures. Notably, most prior work trained only a single network instance per architecture, introducing potential confounds from random weight initialization. Mehrer et al.[53] showed that individual differences across trained networks can be substantial, calling into question the validity of drawing mechanistic insights from single-model comparisons.

Thus, critical evidence on the role of top-down feedback connections in recurrent vision models is still lacking. To address this gap, we trained ConvRNNs on image classification, endowing models with identical feedforward and lateral recurrent connections but differing in the presence or absence of top-down feedback. We then evaluated these models in their speed-accuracy trade-off and sensory robustness in out-of-distribution (OOD) settings using Gaussian noise perturbations and adversarial attacks[54]. Importantly, we matched the number of learnable parameters between models with and without top-down feedback projections, to ensure a fair comparison, and trained multiple model instances per model class.

We tested the core hypothesis that top-down feedback improves image classification especially in OOD regimes. Crucially, we also explored the hypothesis that a putative advantage only emerges under specific training conditions that are commonly faced by biological systems. Specifically, we examined the role of intrinsic neural stochasticity, simulated by injecting dropout[55] during training, and how it interacts with feedback projection.

Adding neural noise allowed us both, to improve biological realism of our simulations, mimicking stochastic synaptic noise, and to disentangle its computational role in regularizing against overfitting compared to the effect of the top-down feedback. To evaluate these hypotheses, we also performed representational similarity analysis, intrinsic dimensionality estimation of latent manifolds, and decoding class information across different unit population sizes.

## Results

**Model setting and training**

We trained ConvRNN models for image classification[42,44], consisting of 3 convolutional layers and a readout layer for classification (Fig. 1A). Models received an image as input, processed it recursively for 10 time steps, and provided a class prediction for each time step that encouraged them to be accurate as fast as possible[56]. We compared ConvRNN with feedforward information flow (FF), relying only on bottom-up and lateral connections, and ConvRNN with additional feedback projections (FB) from the top layer to the bottom one, thus having an additional top-down informational stream. FF and FB models were matched in the total number of learnable parameters. FB ConvRNN models integrated bottom-up and top-down pathways by concatenating the feature maps from both streams and then applying a convolutional layer (Fig. 1B).

Both model types were trained with (dropout on, $p = 0.5$) or without dropout (dropout off, $p = 0$), to simulate the effect of intrinsic stochasticity observed in biological systems[55]. During training, test accuracy (measured on an in-distribution test set) reached about 70% after 50 epochs for all model types (averaging across all time steps), with FF models having slightly higher accuracy than FB models, regardless of the dropout being present or not (Fig. 1C). Similarly, the entropy of the predictive distribution (final layer after softmax) was slightly lower for FF compared to FB across dropout model types (Fig. 1D).

**Speed-accuracy trade-off in recurrent vision models**

We thus investigated the trade-off between speed and accuracy in these models by selecting entropy thresholds to decide when to terminate the recurrent image processing and make a prediction[56]. Plotting reaction times (RT) and accuracy as a function of entropy threshold revealed clear differences between model types (Fig. 1E). We then optimized the entropy threshold to obtain a measure of the speed-accuracy trade-off in each model, by selecting the one that yielded the highest accuracy with the lowest reaction time, equally weighting these metrics of performance. This procedure involved halting the decision process at the first time step where the entropy of the predictive distribution dropped below the threshold, repeating this process for each candidate entropy threshold (see methods). When evaluating the outcome of this optimization procedure, we found that model instances were clustered by the factorial inclusion of dropout and feedback information (Fig. 1F). These entropy thresholds were generally higher in FB compared to FF regardless of dropout inclusion (Fig. 1G). Remarkably, FB with dropout entropy thresholds were higher than in any other model type, suggesting that these models followed a more "cautious" strategy in the decision process.

We aggregated accuracy and RT in a single score to evaluate the speed-accuracy trade-off. We found that FB ConvRNN models exhibited a better speed-accuracy trade-off compared to FF models regardless of dropout inclusion (Fig. 1H). Crucially, FB with dropout models outperformed all the other models in trading-off speed and accuracy, losing only ~10% of accuracy compared to FF without dropout but being around twice as fast. These findings showed how the combination of top-down feedback and stochasticity substantially reshaped model behaviour, endowing recurrent vision models with a markedly superior speed–accuracy trade-off compared to models lacking either component.

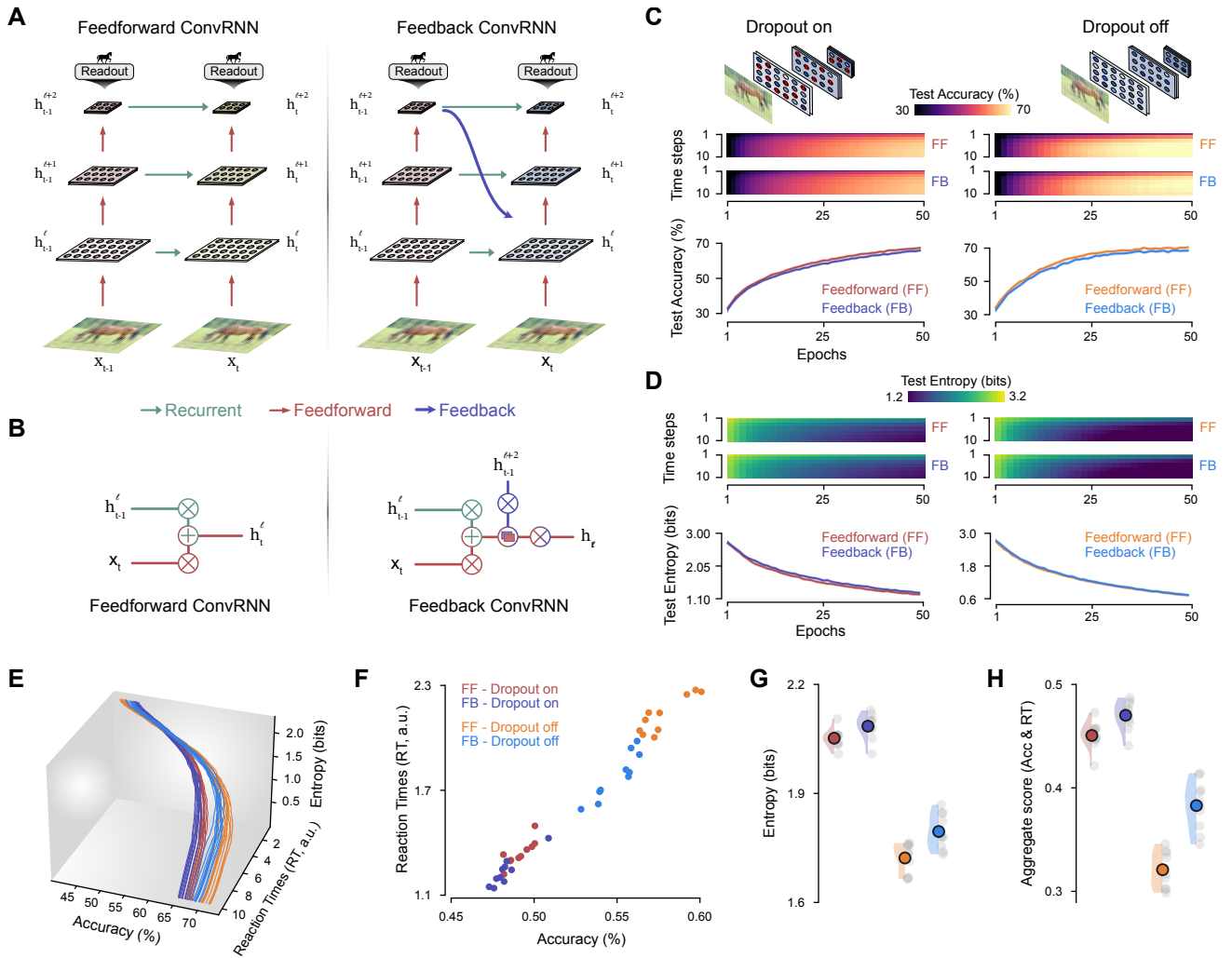

**Figure 1 | Convolutional RNN with feedback from top layers. A.** Illustration of the architecture with recurrent, feedforward and feedback connections. **B.** Illustration of the computational mechanisms for integrating the bottom-up and top-down stream, by concatenating the feature maps and applying a convolutional layer for information integration. **C.** Test accuracy during training across epochs (bottom) and epochs by time (top) for Feedforward-only (FF) and Feedback (FB) ConvRNN with or without dropout. **D.** Entropy of the predictive distribution (final layer after softmax) in the test set during training across epochs (bottom) and epochs by time (top) for FF and FB ConvRNN with or without dropout. **E.** 3D Lineplot showing the accuracy and reaction times (RT) as a function of the entropy threshold. **F.** Scatterplot depicts the accuracy and RT for each model instance by optimizing the entropy threshold in order to maximize accuracy and minimize RT. Raincloud plots showing the values of the optimized entropy thresholds (**G.**) and the aggregate score of accuracy and RT (**H.**). Each dot represents a model instance, while the circle overlapping the density plot represent the mean of the distribution

**Combining feedback and neural noise yields sensory robustness**

After establishing model behaviour on in-distribution test data, we next tested our key hypothesis, i.e. that feedback projections improve sensory robustness in OOD samples, by perturbing the incoming data with different types of sensory noise. First, we added Gaussian noise to the test images, sampling independently at each time step, with a varying percentage ($\alpha$) of corrupted pixels (Fig. 2A). Surprisingly, we found that noise-free models (no dropout) with feedback projections were not more accurate than feedforward models as noise perturbation progressed. FF outperformed FB models both in terms of peak accuracy and persistence across time steps. In stark contrast, we found that models trained with intrinsic neural variability (dropout on) exhibited the opposite pattern, with FB models outperforming FF models as noise perturbation progressed, both in terms of peak accuracy and persistence across time steps. Crucially, models with top-down feedback and dropout outperformed all other models in terms of sensory robustness, i.e. exhibited higher peak accuracy and more sustained performance under escalating noise perturbations.

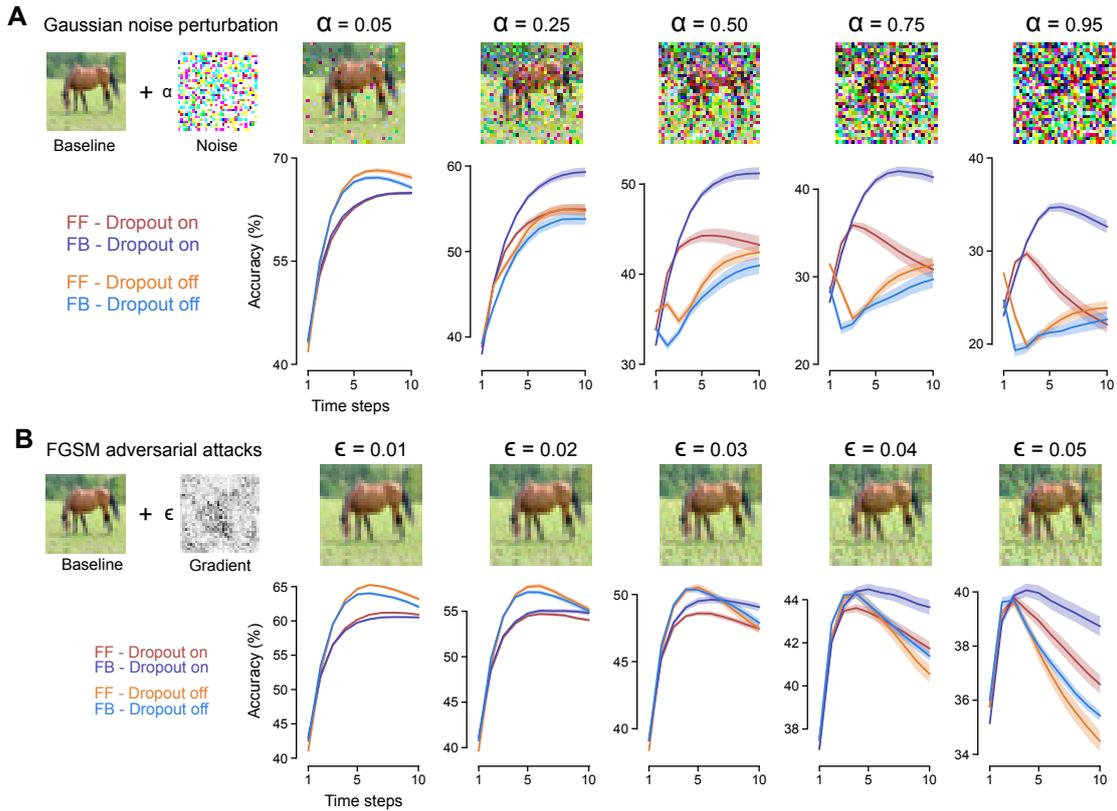

**Figure 2 | Sensory robustness of ConvRNN with feedback and dropout. A.** On the left, illustration of the method for Gaussian noise perturbation parametrized by $\alpha$. On the right, lineplots showing the accuracy as a function of time steps on the test set with varying levels of $\alpha$ for Feedback-only (FF) and Feedback (FB) ConvRNN with or without dropout. Shaded areas represent standard error of the mean (SEM) across model instances. **B.** On the left, illustration of the method for Fast Gradient Sign Method (FGSM) adversarial attacks parametrized by $\epsilon$. On the right, lineplots showing the accuracy as a function of time steps on the test set with varying levels of $\epsilon$ for FF and FB ConvRNN with or without dropout. Shaded areas represent SEM across model instances.

Next, we tested whether these results generalize to different OOD regimes. We performed adversarial attacks via the Fast Gradient Sign Method (FGSM)[54] on test images and parametrized the level of injected noise with an $\epsilon$ parameter (Fig. 2B). Here, we observed that, in models trained in noise-free conditions, FB models showed higher peak and persistent accuracy compared to FF models on high levels of injected noise. Remarkably, models trained with dropout that could rely on top-down information (FB) outperformed their FF counterparts — and all other models — in terms of peak accuracy and persistence across time steps as the adversarial attacks became stronger. Together, these results suggested that the combination of dropout-like neural noise and top-down information made the ConvRNN models substantially more robust to noise perturbation than any other model type which have either these components in isolation or none of them. In the rest of the analyses, we focused on investigating what makes the combination of feedback information and dropout so special.

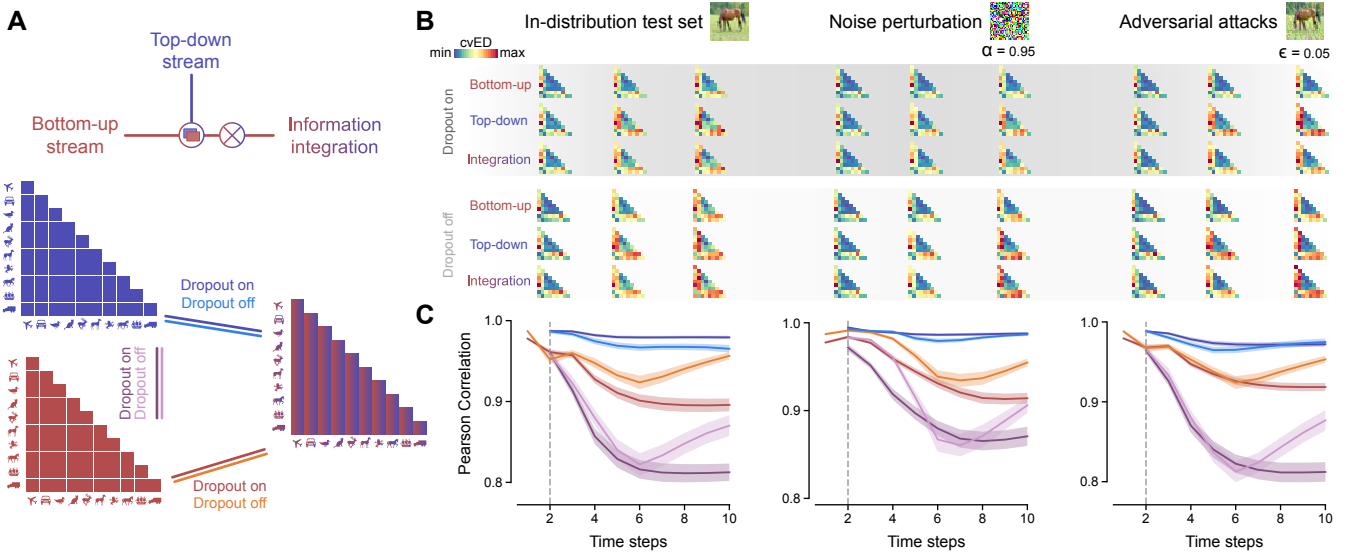

**Figure 3 | Representational similarity analysis showing information flow across streams. A.** Graphical illustration of the Representational Similarity Analysis (RSA) applied to the bottom-up and top-down stream in the ConvRNN models with feedback information. We compared the representational dissimilarity matrices (RDM) encoding pairwise distances between image classes centroids computed on the layer activations coming from the first layer (pre-integration), the feedback projection and the integration layer (post-integration). We correlated these RDMs between these informational streams to reveal representational alignment as a function of time steps. **B.** RDMs plotted at time step 1, 5, and 10 for bottom-up, top-down and integration streams in both FB ConvRNN with or without dropout and across the in-distribution test set, the test set with noise perturbation ($\alpha$=0.95) and the test set with adversarial attacks ($\epsilon$=0.05). Each RDM is scaled by its minimum and maximum cross-validated Euclidean distance (cvED) value. **C.** Lineplots showing the Pearson correlation between RDMs as a function of time steps. The grey dashed vertical line indicates the start of the first feedback projection. Note that the correlation between top-down and information integration (dark and light blue) starts from time step 2 because before the feedback information is not available. Shaded areas represent SEM across model instances.

**Representational similarity analysis to trace information flow**

We first applied Representational Similarity Analysis (RSA) to examine the representational geometry of the bottom-up (BU, i.e. the first layer pre-integration), and top-down (TD) information streams (Fig. 3A) in the first layer, focusing only on models equipped with feedback connections. Specifically, we computed representational dissimilarity matrices (RDMs) for each stream (Fig. 3B), as well as for the combined representation after integration (IT). By comparing these RDMs, we were able to quantify the relative contributions of feedforward versus feedback signals and assess how dropout alters that balance.

We found that these models substantially relied on top-down feedback information across all data regimes, from the in-distribution test set to noise perturbation and adversarial attack OOD data (Fig. 3C). This result was evidenced by the fact that the correlation between TD and IT RDMs was higher than the correlation between BU and IT RDMs across all time steps. Notably, we observed that dropout amplified this effect, since TD-IT correlations were generally higher for dropout models than their non-dropout counterparts across timesteps and data regimes. Moreover, dropout altered the late-stage dynamics (time steps ≥ 6), which showed divergent patterns of correlation among BU, TD and IT streams. Here, the BU-TD and BU-IT correlations suddenly tended to increase in models trained without dropout, while dropout models' representational dynamics remained stable. Together, these findings demonstrate that feedback profoundly shapes ConvRNN representational geometry, and that this top-down signal combined with intrinsic neural stochasticity stabilizes representational dynamics by reinforcing top-down contributions.

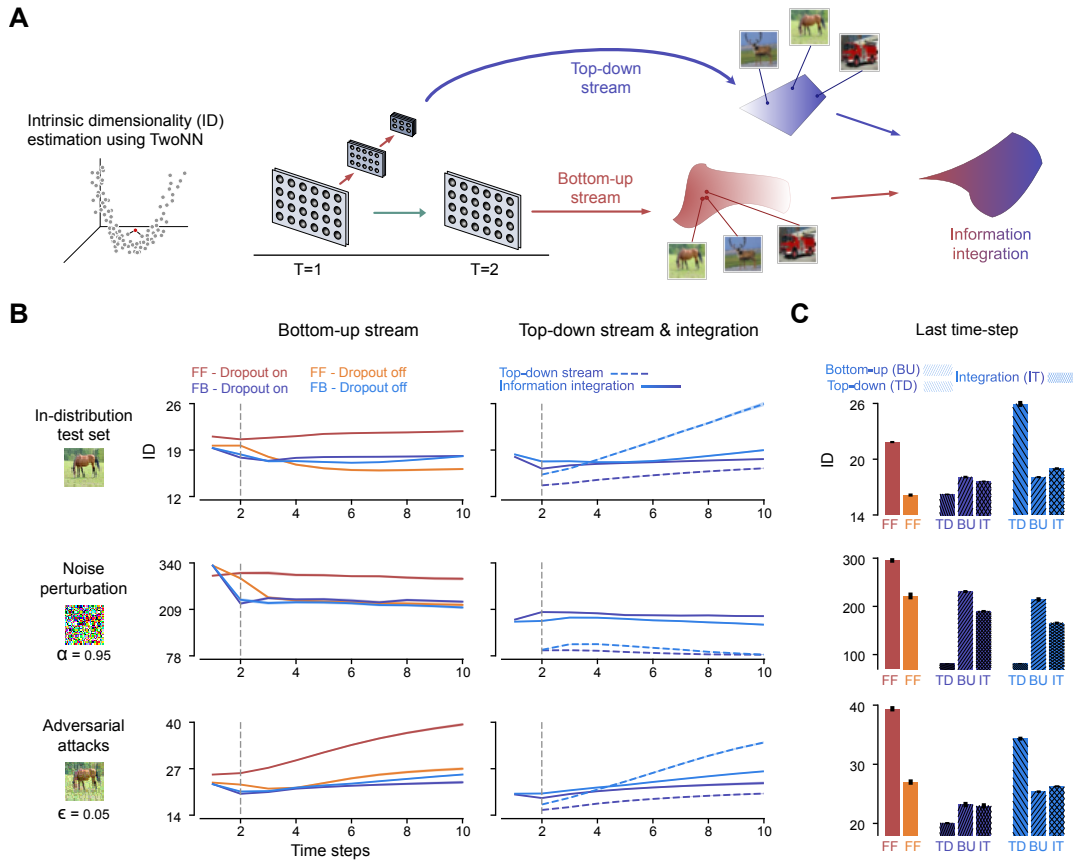

**Figure 4 | Investigating the latent manifold of informational streams. A.** On the left, graphical illustration of the intrinsic dimensionality (ID) estimation procedure using the TwoNN estimator. On the right, illustration of the general procedure used to test the hypothesis on the dimensionality of the latent manifold underlying the layer activations across bottom-up, top-down and integration streams. We tested whether the feedback projection lies on a low-dimensional manifold that constrains the bottom-up information after the integration. **B.** Lineplots showing the ID values as a function of time steps for Feedback-only (FF) and Feedback (FB) ConvRNN with or without dropout in the first column (bottom-up stream), while in the second stream the top-down (dashed) and information integration (solid) ID values only for FB ConvRNN models. These results are presented for the in-distribution test set, the test set with noise perturbation ($\alpha$=0.95) and the test set with adversarial attacks ($\epsilon$=0.05) across the rows. The grey dashed vertical line indicates the start of the first feedback projection. Note that the ID estimation of the top-down stream (dashed dark and light blue in the second column) starts from time step 2 because before the feedback information is not available. Shaded areas represent SEM across model instances. **C.** Barplots showing the same ID values in **B.** but only focused on the last time point. Errorbars indicate SEM across model instances.

### Feedback and dropout reduce manifold's intrinsic dimensionality

Next, we hypothesized that the feedback projection occupies a low-dimensional manifold that, upon integration, constrains the richer bottom-up activity. To test this hypothesis, we applied the TwoNN[57] intrinsic dimensionality (ID) estimator to first layer activations from the bottom-up, top-down, and post-integration streams of both FF and FB ConvRNNs, with and without dropout (Fig. 4A).

In the BU stream, we found that FF models with dropout had a markedly higher ID across all time steps and data regimes compared to all the other model classes (Fig. 4B, left column). This likely reflects a dropout effect on broadening representations rather than collapsing them, spreading information more evenly across a higher-dimensional manifold. Moreover, although in the absence of dropout FB models exhibited a marginal decrease in ID compared to its FF counterpart under OOD regimes, the introduction of dropout amplified this gap. FB with dropout maintained a significantly lower ID than its FF counterpart across all data regimes, indicating that feedback projections synergize with simulated neural noise to yield a more compact, low-dimensional manifold underlying layers' activations.

In the TD stream, as soon as feedback became available (Fig. 4B, middle column, time step 2 onward), FB models trained with dropout showed a strikingly lower ID compared to FB models without dropout across all data regimes and time steps. Following integration (IT), models combining feedback projections with dropout exhibited generally lower ID across time steps than their non-dropout counterparts on in-distribution and adversarial attack data regimes (Fig. 4B, middle column), while on noise perturbed samples we found the opposite pattern. We then focussed on the final time step (Fig. 4C), to directly compared the ID of BU and IT to quantify how much the TD pathway influenced the latent manifold's intrinsic dimensionality of the post-integration (IT) layer activations.

We found that post-integration ID was lower than pre-integration ID only in the presence of both feedback and dropout across all data regimes (for models without dropout this was the case only on noise perturbed samples), confirming that combining top-down feedback signals with dropout-like noise is critical for constraining ConvRNN model activity onto a low-dimensional manifold.

**Decoding object class information scaling across population sizes**

Finally, we asked how feedback projections and simulated neural variability shape both the decodability and population-level stability of object class representations. Here, population-level stability refers to the degree to which information is distributed across many units rather than concentrated in a few. To assess this, we randomly sampled sub-populations of 1-200 units from the first layer in the bottom-up, top-down and integrated informational streams, and measured the mutual information between their activations and class labels using Gaussian Copula Mutual Information[58] (GCMI, Fig. 5A). We operationalized representational stability as the slope of decoding performance across population sizes, based on the idea that more distributed representations yield stronger scaling with larger populations. We focused our decoding analysis on the final time step, where behavioural differences between models under noise and adversarial perturbations were maximal.

In the bottom-up stream (Fig. 5B, left), FB ConvRNNs showed consistently higher decoding performance at each population size than FF models across all regimes, as expected given that FB first layer activations already incorporate top-down signals in the final time step. At this stage of the network hierarchy, no clear benefit of dropout was observed on FB models, as models with dropout exhibited less GCMI on in-distribution data regime and only modest higher decoding performance at higher scale on OOD samples compared to their non-dropout counterparts.

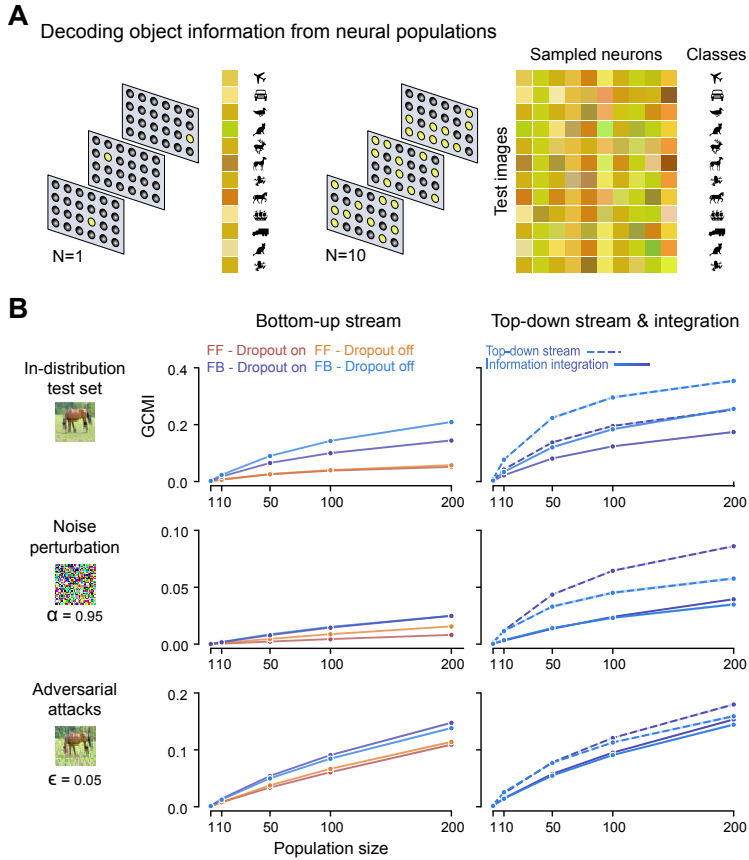

**Figure 5 | Decoding object information from artificial neuronal populations of increasing size. A.** Graphical illustration of the method used to decode object information from neural populations. We sampled sets of artificials neurons (with numerosity ranging from 1 to 200 from the last time point) and used their activations to predict image classes using the Gaussian Copula Mutual Information (GCMI). **B.** Lineplots showing the GCMI values as a function of neural population size for Feedback-only (FF) and Feedback (FB) ConvRNN with or without dropout in the first column (bottom-up stream), while in the second stream the top-down (dashed) and information integration (solid) GCMI values only for FB ConvRNN models. These results are presented for the in-distribution test set, the test set with noise perturbation ($\alpha$=0.95) and the test set with adversarial attacks ($\epsilon$=0.05) across the rows. Shaded areas represent SEM across model instances.

Strikingly, in both top-down and integrated information streams (Fig. 5B, right), FB models with dropout exhibited higher decodability at each population size than their non-dropout counterparts in both OOD data regimes as well as consistent higher population-level stability (dropout on/off; TD, noisy slope = $4.1/2.7 \times 10^{-4}$, adversarial slope = $8.7/7.5 \times 10^{-4}$; IT, noisy slope = $2.0/1.7 \times 10^{-4}$, adversarial slope = $7.5/7.1 \times 10^{-4}$), while showing the opposite pattern on in-distribution samples, suggesting that dropout bolsters robustness to distributional shifts while tempering encoding fidelity under standard conditions. In sum, these findings suggest that, although dropout introduces variability at the single-unit level, its combination with top-down feedback stabilizes representations at the population level while preventing overfitting, achieving an optimal balance between stability and generalization.

## Discussion

Here, we showed how the synergistic interplay between top-down feedback and dropout-like intrinsic neural variability endows recurrent vision models with a remarkable sensory robustness to out-of-distribution samples. Contrary to the hypothesis that top-down feedback alone would suffice for sensory robustness, we found no evidence that feedback by itself confers such robustness. Rather, our results show that only models endowed with both top-down feedback and intrinsic neural variability exhibited superior speed–accuracy trade-off and markedly enhanced resilience to noise perturbations and adversarial attacks. These results highlight the relevance of addressing the relative advantages of feedforward versus feedback processing using perceptual tasks that share key features (noise perturbation, neural variability) as those faced by biological organisms.

Our findings underscore a broader disconnect between biological and current artificial vision systems. While extensive feedback pathways are ubiquitous in cortex, most state-of-the-art Artificial Intelligence (AI) models remain strictly feedforward. Early attempts to harness feedback in machine learning models date back to Deep Boltzmann Machines and the wake-sleep algorithm[59–62], which used top-down generative models during both wake and sleep phases to shape hierarchical representations. As one striking example, these unsupervised networks developed a "visual number sense" capable of supporting numerical judgments, despite never being explicitly trained to count[63,64]. Recent neuroAI efforts have begun to revisit recurrent and feedback motifs, yet it often remains unclear whether reported gains stem from true top-down projections, from local recurrence, or simply from changes in network capacity or parameter count. Although our results on the limited advantage of feedback projections may seem in contrast with this recent literature, our study introduces several critical distinctions. Our model architectures differ from feedforward networks solely by the inclusion of feedback, total parameter counts are matched and multiple instances per model class were trained and evaluated. Crucially, none of the prior studies applied dropout or comparable regularization during training[43–52]. By showing that feedback delivers substantial robustness gains only when paired with dropout-like neural noise, we offer a parsimonious explanation for the limited evidence of top-down induced sensory robustness reported in those earlier works[42,44]. Notably, one prior study[65] did incorporate dropout into a feedforward–feedback loop architecture and reported benefits of feedback, but it relied on a complex, self-attention–based model and demonstrated these effects on tasks unrelated to image classification.

Interestingly, the benefits we observed from combining dropout with feedback echo a well-documented phenomenon in sensory neuroscience, known as stochastic resonance[66], whereby the addition of a certain level of noise can paradoxically enhance signal detection and information processing in sensory systems. Classic studies have shown that subthreshold visual or somatosensory stimuli become more readily detectable when embedded in optimal levels of neural or external noise, a counterintuitive effect observed across species and modalities[67,68]. In the context of ConvRNNs, dropout-like neural stochasticity may play a similar role (albeit at a different timescale), temporarily perturbing unit activations in a way that sharpens feature selectivity and prevents the model from becoming trapped in suboptimal representational regimes. Notably, a recent study highlighted how the combination of predictive coding as top-down feedback and stochastic resonance as bottom-up process may provide explanatory power for the emergence of auditory phantom perception[69]. Moreover, another recent study has also demonstrated that injecting such intrinsic variability into deep vision models enhances their resilience to adversarial attacks[70], in line with our findings. Crucially, we extended these results by showing that the combination with top-down feedback produces a markedly more powerful mechanism for sensory robustness.

Another interesting link of our results to the neuroscientific literature is the phenomenon of representational drift, where individual neurons' tuning can wander substantially over time yet downstream readouts remain stable at the neural population level[71–73]. Aitken et al.[74] showed that continual learning in artificial neural networks, when coupled with dropout but no other types of noise, reproduces the gradual turnover of single-cell tuning seen in mouse visual cortex, yet population-level decoding remained stable. These mechanisms accord well with our findings and provide an account for how the brain build resilient cortical maps[75], with stochastic perturbations introducing flexibility and preventing overfitting, and top-down feedback restoring representational stability at the population level.

This view of top-down feedback as a stabilizing scaffold is closely related to, but distinct from, theories of predictive coding, which posit that feedback conveys top-down predictions that are compared with bottom-up sensory inputs to compute prediction errors[15,17–19,21–23]. Minimizing such prediction errors aligns sensory representations with internal models of the world and reduce the computational load on downstream areas[15,76]. While both perspectives emphasize the role of feedback in maintaining internal coherence, predictive coding primarily frames feedback as a way to convey predictions that explain away predictable stimuli[18], rather than explicitly as a stabilizing force. However, these functions are not mutually exclusive: by suppressing predictable or redundant activity, especially in environments where inputs are noisy or ambiguous, the predictive model itself may act as a stabilizing prior, guiding the network toward consistent interpretations (or posterior beliefs) across time and avoiding that the network dynamics simply track noise[6,77].

Furthermore, the dropout mechanism used in this study has formal relations to Bayesian model reduction[78], which is used to prune unnecessary parameters and reduce the complexity of predictive coding models – and arguable acts as another stabilizing force, by avoiding overfitting. Notably, here we avoided hard-wiring a specific predictive coding network architecture endowed with prediction and prediction error units[22,47,51,79] and instead took an agnostic approach[65], allowing a learnable convolutional integration layer to discover the most effective way to merge bottom-up and top-down streams. Future work could directly compare alternative predictive coding integration mechanisms to tease apart how predictive feedback contributes to population-level representational stabilization, as well as evaluating these effects in standard predictive coding networks[22,80–82] equipped with simulated neural variability. Moreover, future work could also investigate if the precision-control mechanisms[83] used in predictive coding networks to contextually modulate the relative contributions of feedforward and feedforward streams are also effective in our ConvRNNs architecture.

Our results show how the interplay between feedback and neural variability constrains model activity onto a low-dimensional manifold. This is in line with recent theoretical and experimental work showing the benefits of having a low-dimensional manifold for sensory robustness in both biological and artificial systems[84,85]. In biological systems, brain responses during visual tasks reliably collapse onto only a handful of latent dimensions, yielding smooth, noise-tolerant trajectories and stable manifold topology[86–91]. In parallel, theoretical deep learning research has revealed how reductions in manifold radius and dimension underlie improved classification capacity and compact latent manifolds yield better resistance to adversarial attacks[92–94].

Furthermore, our findings show how feedback projections and neural stochasticity improve the decodability of object class information in subsequent layer representations, by substantially shaping the bottom-up stream towards the linearly separability of the image classes. This accord well with recent findings on the computational mechanisms adopted by deep neural networks to solve real-world tasks like image classification, which shows how the model gradually reformats the data representations through the hierarchical layer structure into increasingly linearly separable feature spaces, culminating in a final representation that can be readout with simple classifiers[95–98].

The efficiency and robustness of recurrent networks combining feedback and dropout are also in line with variational accounts. In particular, the free energy principle[76] treats classification as probabilistic inference under a generative model. In these formulations, rooted in the ideas of Helmholtz machines and hierarchical predictive coding[99–101], the network's architecture encodes priors in its deeper layers, which are then dynamically integrated with bottom-up sensory evidence. Crucially, only architectures with feedback descending connections allow the combination of empirical priors from the deepest layers with likelihood messages from the input (sensory) layers. This key architectural feature, a hallmark of biological visual cortical hierarchies[102–104], enables the model to flexibly integrate global context with local inputs beyond lateral conections. In variational schemes, the free energy serves as a tractable bound on the model evidence (ELBO)[105], which can be decomposed into an accuracy and complexity term[106]. This means that complexity can be understood as the degrees of freedom that are used in providing an accurate classification. Maximising the model evidence, therefore, implicitly penalize model complexity to provide a parsimonious and accurate representation of the training data, which generalises to OOD test data.

We highlight two useful insights from this perspective. The first is that minimising complexity appeals to exactly the same principles found in minimum description length schemes based upon algorithmic complexity[107–109]. In turn, this can be viewed in terms of compression[110]. In other words, if the test data can be compressed or represented on a low dimensional manifold, then model evidence will be higher leading to better generalisation and robustness to OOD test data. Effectively, this is exactly what we have found. The representational encoding of the images in a low dimensional space speaks to compression and the rationale for all information bottleneck architectures[111–114]. Moreover, it may explain why dropout was necessary to realize this compression. As alluded to above, Bayesian model reduction leverages the dropout of model parameters to find models with greater evidence (i.e., finding models in which the complexity can be decreased by dropout). The objective function used in our RNN training was limited to classification accuracy, thereby requiring dropout to regularize model complexity. It is noteworthy that when complexity is regularised, the feedback schemes compress the input data more efficiently, leading to more robust and efficient inference[115].

These findings carry several implications for both neuroscience and machine learning. From a neuroscientific perspective, our results suggest a fundamental stabilizing role for cortical feedback, counteracting intrinsic neural variability to confine neural activity onto low-dimensional manifolds. This view complements and extends classic modulatory and predictive-coding accounts of top-down feedback projections[3,4,15,76,116], highlighting its active role in preserving representational stability at the population level despite intrinsic neural stochasticity. From a machine-learning standpoint, our findings pave the way for biologically-inspired vision architectures that move beyond purely feedforward[25–27] or locally recurrent[41,44] designs by incorporating explicit top-down pathways alongside stochastic regularization. Such models not only promise enhanced sensory robustness, but also improved interpretability through clearly delineated feedback control and new goal-directed encoding capacities, providing a dynamic interface between high-level objectives and visual processing[117].

In conclusion, our findings uncover a dual mechanism for sensory robustness, in which neural stochasticity prevents unit-level co-adaptation by decorrelating activations, thereby reducing overfitting albeit at the cost of more chaotic dynamics, while top-down feedback leverages high-level information to constrain network activity on compact low-dimensional manifolds and maintains representational stability at the population-level.

## Methods

**Model architecture and training**

We trained convolutional recurrent neural networks (ConvRNN) that recursively processes a static image $x \in \mathbb{R}^{H \times W \times C}$ at each step $t \in \{1, ..., T\}$, with all pixel intensities scaled to lie in the interval [0,1]. The base architecture consists of a stack of $L$ downsampling convolutional layers compressing the spatial resolution to $h \times w$ with $h = H/2^L$, $w = W/2^L$, by setting the stride to 2. Let $h_t^\ell \in \mathbb{R}^{h_\ell \times w_\ell \times C_\ell}$ be the feature map hidden state for layer $\ell$ and recursive timestep $t$, with $h_t^1 = x$ and $C_\ell$ being the number of feature map channels, the bottom-up feedforward process is described as:

$$h_t^\ell = \sigma(\mathcal{W}^\ell * h_t^{\ell-1} + \mathcal{U}^\ell * h_{t-1}^\ell) \tag{1}$$

where $\sigma(\cdot)$ denotes the pointwise tanh activation function, $*$ represents the 2D convolution operator, $\mathcal{W}^\ell \in \mathbb{R}^{k \times k \times C_{\ell-1} \times C_\ell}$ and $\mathcal{U}^\ell \in \mathbb{R}^{k \times k \times C_\ell \times C_\ell}$ are the kernels of the convolutional layers for the input-to-hidden and the hidden-to-hidden state, respectively, and $k$ is the kernel size. At the first time step $t$=1, the previous hidden state was initialized as zeros.

ConvRNN models with top-down modulation were implemented with a long-range feedback connection from the last layer $L$ to the first one $\ell$=1. To match the dimensionality of the first layer, we upsampled the hidden state of the last layer $h_{t-1}^L$ with two transpose 2D convolutions with stride set to 2. The integration mechanisms for combining bottom-up and top-down streams consisted of concatenating their hidden states along the channel dimension and then process it by a convolutional layer:

$$h_t^{BU} = \sigma(\mathcal{W}^1 * x + \mathcal{U}^1 * h_{t-1}^1) \tag{2}$$

$$h_t^{TD} = Upsample(h_{t-1}^L) \tag{3}$$

$$h_t^1 = \sigma(V * [h_t^{BU} \parallel h_t^{TD}]) \tag{4}$$

where $Upsample(\cdot)$ denotes the feedback upsampling operation, $\parallel$ indicates the channel-wise concatenation operation and $V \in \mathbb{R}^{k \times k \times (C_1 + C_3) \times C_1}$ is the kernel for the convolutional layer that integrate the feedforward and feedback signals. For models with dropout enabled during training, we randomly silencing units with a probability $p$ after the tanh nonlinearity stage at each layer and time step $h_t^\ell$. Notably, these masked hidden state activations were passed to subsequent processing as lateral recurrent and feedback connections. At each time step $t$, the final hidden state $h_t^L$ is first reduced via global average pooling, then passed through a linear readout layer to produce logits $z_t \in \mathbb{R}^Q$ for a $Q$-class classification problem and finally a softmax function is applied to yield the predictive probability distribution. Notably, since the logits $z_t$ are generated at each recurrent step, the model can perform dynamic inference, potentially trading off accuracy and speed.

All models comprised of 3 convolutional layers. The number of feature maps $C_\ell$ in models with top-down feedback connections was set to 64, 128 and 256. This resulted in a total number of parameters equal to 1 590 602, with about 27.8% of connections deputed to the top-down pathway (368 832 parameters for the upsampler and 73 792 for the integration convolutional layer). To match the number of parameters for the feedforward ConvRNN models, we set the number of feature maps to 80, 160 and 288, resulting in a total number of learnable parameters equal to 1 570 522. The kernel size $k$ was set to 3 for all convolutional and transpose convolutional layers and the dropout probability $p$ was set to 0.5 for models with dropout enabled.

The models were trained by minimizing the cumulative cross-entropy loss over all $T$ steps, thereby encouraging the system to be accurate as fast as possible. Models were trained for 50 epochs, with a batch size of 1024 and unrolling for $T$=10 time steps. Optimization was performed with the Adam optimizer[118] (learning rate γ=0.0003, $β_1$=0.9, $β_2$=0.999), with gradient norm clipping to stabilize training by setting the maximum norm to 1. All parameters were initialized with Xavier (Glorot) initialization. Importantly, we instantiated 10 independent replicas for each combination of feedback (present vs. absent) and dropout (enabled vs. disabled), reinitializing parameters for each run.

The models were trained on CIFAR-10[119], a standard image classification benchmark composed of 50 000 training and 10 000 test RGB images (32×32 pixel resolution), spanning 10 object categories. We retain the canonical train/test split and apply no data augmentation or corruption during training, ensuring that any robustness advantage observed on out-of-distribution test samples stems solely from the recurrent dynamics rather than from prior exposure to similar image perturbations.

**Model evaluation on in-distribution and out-of-distribution data regimes**

We evaluated these models starting from investigating the speed-accuracy trade-off in the in-distribution test set. To address this, we computed the Shannon entropy of the predictive probability distribution over classes for each test sample and time step. We then defined a dense grid of 100 candidate entropy thresholds covering the full range observed across the test set. For each threshold, we measured the classification accuracy and average reaction times (RT), normalized by the maximum amount of time steps $T$, by halting the decision process at the first time step where the entropy of the predictive distribution dropped below the threshold.

We then selected the entropy threshold that maximized the accuracy and simultaneously minimized RT, by selecting the threshold with the highest value computed as accuracy minus the normalized RT. We also computed an aggregated score of performance, by averaging accuracy and RT scores optimized by the entropy thresholds for each model instance. RT scores were obtained by mapping the raw RT to 1 - RT/$RT_{max}$, where $RT_{max}$ is the maximum observed reaction time across all model instances. Notably, normalizing this RT score by $T$ instead of $RT_{max}$ yielded the same qualitative ranking of model classes. Thus, the higher this aggregated score is, the better is the performance, signalling both high accuracy and fast reaction times.

Next, we evaluated these models on out-of-distribution (OOD) samples using two different types of sensory noise. First, we perturbed test images by adding zero-mean Gaussian noise, with a standard deviation of 0.8, independently sampled across test samples and time steps. In other words, at each recurrent step the model was fed a distinct noisy variant of the same image. After the noise addition, pixel intensities were clipping back into the original pixel range. The percentage of corrupted pixels was controlled by an $α$ parameter and five levels were evaluated, i.e. 5%, 25%, 50%, 75%, and 95%.

Second, we perturbed test images using adversarial attacks via the Fast Gradient Sign Method (FGSM)[54]. To generate adversarial examples, we trained a small two-layer CNN on the train set for 20 epochs, reaching a test accuracy of 71.26%, and used this model as our "attack model". For each clean test image and its class label, we passed this image to the attack model and computed the gradient of the cross-entropy loss with respect to its input pixels, took the element-wise sign of that gradient to get a per-pixel direction of steepest increase, scaled it by the parameter ϵ and added it to the image. We then clipped the resulting pixel values back into the valid [0,1] range to produce the adversarial example. Notably, here at each time step the model received the same corrupted image. We used five attack strngth levels by setting ϵ to 0.01, 0.02, 0.03, 0.04, and 0.05.

**Representational similarity analysis**

We performed Representational Similarity Analysis (RSA) to investigate information flow in the bottom-up (BU) and top-down (TD) information pathways. Focusing on the models endowed with feedback connections, we collected the first layer activations across samples and time steps, with BU as the feedforward activation before the integration $h_t^{BU}$, TD as the feedback activation from the last layer projected to the first one $h_t^{TD}$ and the information integration (IT) as the activation after the integration convolutional layer and tanh nonlinearity $h_t^1$ as in eq. 4.

We applied RSA independently for the three test set, the in-distribution set as well as the OOD sets comprising the Gaussian noisy samples with $\alpha$=95% and the adversarial samples with ϵ=0.05. For each time point, we computed the representational dissimilarity matrices (RDM) by using the cross-validated (squared) Euclidean distance[120] (cvED) across the samples. For a given set of (flattened) activations and class labels, we first partitioned the data into training and test splits via a stratified 5-fold cross-validation scheme. Within each split, we computed the mean activation (centroid) for each class separately in the training fold and in the held-out test fold. For every pair of classes, we computed the difference between their centroids in the training data and the corresponding difference in the test data, then took the dot product of these two difference vectors and divided by the number of features.

This dot product across train and test samples yields an unbiased estimate of the squared Euclidean distance between the two class representations. Repeating this procedure for all class pairs and averaging across folds produces a full RDM for that stream and time point. Finally, we assessed the alignment between bottom-up, top-down, and integrated representations by computing Pearson's correlation coefficients between their vectorized RDMs, doing so separately for the dropout and non-dropout model variants.

**Intrinsic dimensionality estimation**

We estimated intrinsic dimensionality (ID) of the latent manifold underlying the activations in the bottom-up, top-down and integrated streams in the first layer by employing the TwoNN algorithm[57]. Models' activations 3D structure was flattened prior to the ID estimation.

First, we compute the full matrix of pairwise Euclidean distances between all sample vectors in the layer's activation space. For each sample, we then identify its two closest neighbors (excluding itself) and form the ratio of the larger distance to the smaller one. Collecting those ratios across all samples, we build their empirical cumulative distribution by ranking them in ascending order and mapping each rank to its quantile. We then transform the ratios and quantiles via logarithms and fit a straight line through the origin to those transformed points using a least-squares solver. The slope of that line provides the intrinsic dimension estimate.

Intrinsic dimensionality quantifies the true number of degrees of freedom that underlie a collection of high-dimensional data points, i.e. in our case the activation vectors collected in the models' first layer. Whereas the extrinsic dimensionality simply equals the number of features (32×32×80=81 920 for feedforward ConvRNNs in the BU activations; 32×32×64=65 536 for feedback ConvRNNs in BU, TD and IT activations), the intrinsic dimension estimates how many independent factors are really needed to describe the data. A low intrinsic dimension implies that, despite living in a high-dimensional embedding space, the activations lie on a low-dimensional manifold. By contrast, a high intrinsic dimension indicates that activity spans many independent directions, effectively utilizing a large portion of the feature space. We apply this ID estimation algorithm to the BU activations in the feedforward and feedback ConvRNN models (with and without dropout) and the TD and IT activations in the feedback ConvRNNs, separately for each model instance, time step and the in-distribution and OOD test sets.

**Decoding analysis with scaling population sizes**

We performed decoding analysis using the mutual information via Gaussian copula estimator[58] (GCMI) to assess how much class information is present in the bottom-up, top-down and integrated streams of the first layer activity patterns. All analyses were performed in the in-distribution and OOD test sets on the activations at the final recurrent step, when differences across model variants under both clean and perturbed inputs were maximal.

For each stream, we first flattened the spatial activity pattern of the first layer into a high-dimensional feature vector per image. We then assessed decodability by drawing random sub-populations of units of varying sizes. Specifically, we sampled groups of 1, 10, 50, 100, and 200 units. For each population size, we repeated this random selection 100 times, computing the GCMI between the chosen units' joint activity and the ground-truth class labels. Averaging across these repeats yields a robust estimate of how much class information is recoverable from any arbitrary sub-set of that many units.

Beyond overall information content, we were also interested in population-level representational stability, defined as the extent to which that information is broadly distributed rather than tied to a few key units. To capture this, we examined how the decoding performance rises steadily with population size, with the assumption that this continual increase signals a stable population code, one where information is embedded across many units rather than concentrated in just a few. By contrast, if only a small subset of units carries the class signal, the decoding performance curve would climb only until your sample size includes those informative units. Beyond that point, adding more units yields no extra information and the GCMI curve flattens. We therefore operationalized stability as the slope of the decoding performance curve estimated with the ordinary least square method using the population size as independent variable and the GCMI values as dependent variable.

## Data and code availability
All models, code and results will be released upon publication.

## Author contributions
**A.G.**: conceptualization, software, methodology, investigation, formal analysis, validation, data curation, visualization, writing - original draft, writing - review and editing.
**M.D.**: conceptualization, methodology, formal analysis, writing - review and editing.
**K.F.**: conceptualization, writing - review and editing.
**G.P.**: conceptualization, methodology, writing - review and editing.
**M.S.**: conceptualization, supervision, resources, funding acquisition, writing - review and editing.

## Competing interests

The authors declare no competing interests.

## References

1. Lamme, V. A. F. & Roelfsema, P. R. The distinct modes of vision offered by feedforward and recurrent processing. *Trends Neurosci.* **23**, 571–579 (2000).
2. Gilbert, C. D. & Sigman, M. Brain States: Top-Down Influences in Sensory Processing. *Neuron* **54**, 677–696 (2007).
3. Gilbert, C. D. & Li, W. Top-down influences on visual processing. *Nat. Rev. Neurosci.* **14**, 350–363 (2013).
4. Kreiman, G. & Serre, T. Beyond the feedforward sweep: feedback computations in the visual cortex. *Ann. N. Y. Acad. Sci.* **1464**, 222–241 (2020).
5. Dosenbach, N. U., Fair, D. A., Cohen, A. L., Schlaggar, B. L. & Petersen, S. E. A dual-networks architecture of top-down control. *Trends Cogn. Sci.* **12**, 99–105 (2008).
6. Kok, P., Jehee, J. F. & De Lange, F. P. Less is more: expectation sharpens representations in the primary visual cortex. *Neuron* **75**, 265–270 (2012).
7. Pak, A., Ryu, E., Li, C. & Chubykin, A. A. Top-Down Feedback Controls the Cortical Representation of Illusory Contours in Mouse Primary Visual Cortex. *J. Neurosci.* (2019) doi:10.1523/JNEUROSCI.1998-19.2019.
8. Uithol, S., Bryant, K. L., Toni, I. & Mars, R. B. The Anticipatory and Task-Driven Nature of Visual Perception. *Cereb. Cortex* **31**, 5354–5362 (2021).
9. White, A. L., Kay, K. & Yeatman, J. D. High specificity of top-down modulation in word-selective cortex. *J. Vis.* **22**, 3088 (2022).
10. Winding, M. *et al.* The connectome of an insect brain. *Science* **379**, eadd9330 (2023).
11. Greco, A. & Siegel, M. A spatiotemporal style transfer algorithm for dynamic visual stimulus generation. *Nat. Comput. Sci.* (2024) doi:10.1038/s43588-024-00746-w.
12. Greco, A., Rastelli, C., Ubaldi, A. & Riva, G. Immersive exposure to simulated visual hallucinations modulates high-level human cognition. *Conscious. Cogn.* **128**, 103808 (2025).
13. Mumford, D. On the computational architecture of the neocortex. II. The role of cortico-cortical loops. *Biol. Cybern.* **66**, 241–251 (1992).
14. Ullman, S. Sequence seeking and counter streams: a computational model for bidirectional information flow in the visual cortex. *Cereb. Cortex N. Y. N 1991* **5**, 1–11 (1995).
15. Rao, R. P. & Ballard, D. H. Predictive coding in the visual cortex: a functional interpretation of some extra-classical receptive-field effects. *Nat. Neurosci.* **2**, 79–87 (1999).
16. Siegel, M., Kording, K. P. & König, P. Integrating top-down and bottom-up sensory processing by somato-dendritic interactions. *J Comput Neurosci* **8**, 161–73 (2000).
17. Friston, K. A theory of cortical responses. *Philos. Trans. R. Soc. B Biol. Sci.* **360**, 815–836 (2005).
18. Friston, K. *et al.* Active inference and learning. *Neurosci. Biobehav. Rev.* **68**, 862–879 (2016).
19. Pezzulo, G., Parr, T. & Friston, K. The evolution of brain architectures for predictive coding and active inference. *Philos. Trans. R. Soc. B Biol. Sci.* **377**, 20200531 (2022).
20. Rastelli, C., Greco, A., Kenett, Y. N., Finocchiaro, C. & De Pisapia, N. Simulated visual hallucinations in virtual reality enhance cognitive flexibility. *Sci. Rep.* **12**, 4027 (2022).
21. Greco, A., Moser, J., Preissl, H. & Siegel, M. Predictive learning shapes the representational geometry of the human brain. *Nat. Commun.* **15**, 9670 (2024).
22. Millidge, B., Seth, A. & Buckley, C. L. Predictive Coding: a Theoretical and Experimental Review. Preprint at http://arxiv.org/abs/2107.12979 (2022).
23. Greco, A., Rastelli, C., Bonetti, L., Braun, C. & Caria, A. Neural signatures of predictive coding underlying the acquisition of incidental sensory associations. 2025.05.16.654429 Preprint at https://doi.org/10.1101/2025.05.16.654429 (2025).
24. Bonetti, L. *et al.* Shared and modality-specific brain networks underlying predictive coding of temporal sequences. 2025.07.16.665207 Preprint at https://doi.org/10.1101/2025.07.16.665207 (2025).

25. Krizhevsky, A., Sutskever, I. & Hinton, G. E. ImageNet Classification with Deep Convolutional Neural Networks. in *Advances in Neural Information Processing Systems* vol. 25 (Curran Associates, Inc., 2012).
26. LeCun, Y., Bengio, Y. & Hinton, G. Deep learning. *Nature* **521**, 436–444 (2015).
27. Simonyan, K. & Zisserman, A. Very Deep Convolutional Networks for Large-Scale Image Recognition. *ICLR* (2014).
28. He, K., Zhang, X., Ren, S. & Sun, J. Deep residual learning for image recognition. in *Proceedings of the IEEE conference on computer vision and pattern recognition* 770–778 (2016).
29. Khaligh-Razavi, S.-M. & Kriegeskorte, N. Deep Supervised, but Not Unsupervised, Models May Explain IT Cortical Representation. *PLOS Comput. Biol.* **10**, e1003915 (2014).
30. Yamins, D. L. K. *et al.* Performance-optimized hierarchical models predict neural responses in higher visual cortex. *Proc. Natl. Acad. Sci.* **111**, 8619–8624 (2014).
31. Cadieu, C. F. *et al.* Deep neural networks rival the representation of primate IT cortex for core visual object recognition. *PLoS Comput. Biol.* **10**, e1003963 (2014).
32. Güçlü, U. & Van Gerven, M. A. Deep neural networks reveal a gradient in the complexity of neural representations across the ventral stream. *J. Neurosci.* **35**, 10005–10014 (2015).
33. Yamins, D. L. & DiCarlo, J. J. Using goal-driven deep learning models to understand sensory cortex. *Nat. Neurosci.* **19**, 356–365 (2016).
34. Kell, A. J. & McDermott, J. H. Deep neural network models of sensory systems: windows onto the role of task constraints. *Curr. Opin. Neurobiol.* **55**, 121–132 (2019).
35. Richards, B. A. *et al.* A deep learning framework for neuroscience. *Nat. Neurosci.* **22**, 1761–1770 (2019).
36. Zhuang, C. *et al.* Unsupervised neural network models of the ventral visual stream. *Proc. Natl. Acad. Sci.* **118**, e2014196118 (2021).
37. Zador, A. *et al.* Catalyzing next-generation Artificial Intelligence through NeuroAI. *Nat. Commun.* **14**, 1597 (2023).
38. Doerig, A. *et al.* The neuroconnectionist research programme. *Nat. Rev. Neurosci.* **24**, 431–450 (2023).
39. Kanwisher, N., Khosla, M. & Dobs, K. Using artificial neural networks to ask 'why' questions of minds and brains. *Trends Neurosci.* **46**, 240–254 (2023).
40. Khosla, M., Williams, A. H., McDermott, J. & Kanwisher, N. Privileged representational axes in biological and artificial neural networks. 2024.06.20.599957 Preprint at https://doi.org/10.1101/2024.06.20.599957 (2024).
41. Kar, K., Kubilius, J., Schmidt, K., Issa, E. B. & DiCarlo, J. J. Evidence that recurrent circuits are critical to the ventral stream's execution of core object recognition behavior. *Nat. Neurosci.* **22**, 974–983 (2019).
42. Kietzmann, T. C. *et al.* Recurrence is required to capture the representational dynamics of the human visual system. *Proc. Natl. Acad. Sci.* **116**, 21854–21863 (2019).
43. Chung, J., Gulcehre, C., Cho, K. & Bengio, Y. Gated Feedback Recurrent Neural Networks. in *Proceedings of the 32nd International Conference on Machine Learning* 2067–2075 (PMLR, 2015).
44. Spoerer, C. J., McClure, P. & Kriegeskorte, N. Recurrent Convolutional Neural Networks: A Better Model of Biological Object Recognition. *Front. Psychol.* **8**, (2017).
45. Han, K. *et al.* Deep Predictive Coding Network with Local Recurrent Processing for Object Recognition. in *Advances in Neural Information Processing Systems* vol. 31 (Curran Associates, Inc., 2018).
46. Nayebi, A. *et al.* Task-Driven Convolutional Recurrent Models of the Visual System. in *Advances in Neural Information Processing Systems* vol. 31 (Curran Associates, Inc., 2018).
47. Wen, H. *et al.* Deep Predictive Coding Network for Object Recognition. in *Proceedings of the 35th International Conference on Machine Learning* 5266–5275 (PMLR, 2018).
48. Yan, S. *et al.* Recurrent Feedback Improves Feedforward Representations in Deep Neural Networks. Preprint at https://doi.org/10.48550/arXiv.1912.10489 (2019).
49. Ernst, M. R., Triesch, J. & Burwick, T. Recurrent Connections Aid Occluded Object Recognition by Discounting Occluders. in vol. 11729 294–305 (2019).
50. Huang, Y. *et al.* Neural Networks with Recurrent Generative Feedback. in *Advances in Neural Information Processing Systems* vol. 33 535–545 (Curran Associates, Inc., 2020).

51. Choksi, B. *et al.* Predify: Augmenting deep neural networks with brain-inspired predictive coding dynamics. in *Advances in Neural Information Processing Systems* vol. 34 14069–14083 (Curran Associates, Inc., 2021).
52. Lindsay, G. W., Mrsic-Flogel, T. D. & Sahani, M. Bio-inspired neural networks implement different recurrent visual processing strategies than task-trained ones do. 2022.03.07.483196 Preprint at https://doi.org/10.1101/2022.03.07.483196 (2022).
53. Mehrer, J., Spoerer, C. J., Kriegeskorte, N. & Kietzmann, T. C. Individual differences among deep neural network models. *Nat. Commun.* **11**, 5725 (2020).
54. Goodfellow, I. J., Shlens, J. & Szegedy, C. Explaining and Harnessing Adversarial Examples. in *3rd International Conference on Learning Representations, ICLR 2015, San Diego, CA, USA, May 7-9, 2015, Conference Track Proceedings* (eds Bengio, Y. & LeCun, Y.) (2015).
55. Srivastava, N., Hinton, G., Krizhevsky, A., Sutskever, I. & Salakhutdinov, R. Dropout: A Simple Way to Prevent Neural Networks from Overfitting. *J. Mach. Learn. Res.* **15**, 1929–1958 (2014).
56. Spoerer, C. J., Kietzmann, T. C., Mehrer, J., Charest, I. & Kriegeskorte, N. Recurrent neural networks can explain flexible trading of speed and accuracy in biological vision. *PLoS Comput. Biol.* **16**, e1008215 (2020).
57. Facco, E., d'Errico, M., Rodriguez, A. & Laio, A. Estimating the intrinsic dimension of datasets by a minimal neighborhood information. *Sci. Rep.* **7**, 12140 (2017).
58. Ince, R. A. *et al.* A statistical framework for neuroimaging data analysis based on mutual information estimated via a gaussian copula. *Hum. Brain Mapp.* **38**, 1541–1573 (2017).
59. Ackley, D. H., Hinton, G. E. & Sejnowski, T. J. A learning algorithm for boltzmann machines. *Cogn. Sci.* **9**, 147–169 (1985).
60. Hinton, G. E., Dayan, P., Frey, B. J. & Neal, R. M. The 'Wake-Sleep' Algorithm for Unsupervised Neural Networks. *Science* **268**, 1158–1161 (1995).
61. Hinton, G. E., Osindero, S. & Teh, Y.-W. A Fast Learning Algorithm for Deep Belief Nets. *Neural Comput.* **18**, 1527–1554 (2006).
62. Salakhutdinov, R. & Hinton, G. Deep boltzmann machines. in *Artificial intelligence and statistics* 448–455 (PMLR, 2009).
63. Stoianov, I. & Zorzi, M. Emergence of a 'visual number sense' in hierarchical generative models. *Nat. Neurosci.* **15**, 194–196 (2012).
64. Testolin, A., Zou, W. Y. & McClelland, J. L. Numerosity discrimination in deep neural networks: Initial competence, developmental refinement and experience statistics. *Dev. Sci.* **23**, e12940 (2020).
65. Mittal, S. *et al.* Learning to combine top-down and bottom-up signals in recurrent neural networks with attention over modules. in *International Conference on Machine Learning* 6972–6986 (PMLR, 2020).
66. McDonnell, M. D. & Ward, L. M. The benefits of noise in neural systems: bridging theory and experiment. *Nat. Rev. Neurosci.* **12**, 415–425 (2011).
67. Levin, J. E. & Miller, J. P. Broadband neural encoding in the cricket cereal sensory system enhanced by stochastic resonance. *Nature* **380**, 165–168 (1996).
68. Collins, J. J., Imhoff, T. T. & Grigg, P. Noise-enhanced tactile sensation. *Nature* **383**, 770–770 (1996).
69. Schilling, A. *et al.* Predictive coding and stochastic resonance as fundamental principles of auditory phantom perception. *Brain* **146**, 4809–4825 (2023).
70. Dapello, J. *et al.* Neural Population Geometry Reveals the Role of Stochasticity in Robust Perception. in *Advances in Neural Information Processing Systems* vol. 34 15595–15607 (Curran Associates, Inc., 2021).
71. Driscoll, L. N., Pettit, N. L., Minderer, M., Chettih, S. N. & Harvey, C. D. Dynamic Reorganization of Neuronal Activity Patterns in Parietal Cortex. *Cell* **170**, 986-999.e16 (2017).
72. Rule, M. E. & O'Leary, T. Self-healing codes: How stable neural populations can track continually reconfiguring neural representations. *Proc. Natl. Acad. Sci.* **119**, e2106692119 (2022).
73. Ratzon, A., Derdikman, D. & Barak, O. Representational drift as a result of implicit regularization. *eLife* **12**, (2023).
74. Aitken, K., Garrett, M., Olsen, S. & Mihalas, S. The geometry of representational drift in natural and artificial neural networks. *PLOS Comput. Biol.* **18**, e1010716 (2022).
75. Ziv, Y. Resilient cortical maps. *Nat. Neurosci.* **28**, 1368–1369 (2025).

76. Friston, K. The free-energy principle: a unified brain theory? *Nat. Rev. Neurosci.* **11**, 127–138 (2010).
77. Kok, P., Mostert, P. & De Lange, F. P. Prior expectations induce prestimulus sensory templates. *Proc. Natl. Acad. Sci.* **114**, 10473–10478 (2017).
78. Friston, K. J. *et al.* Active Inference, Curiosity and Insight. *Neural Comput.* **29**, 2633–2683 (2017).
79. Hosseini, M. & Maida, A. Hierarchical Predictive Coding Models in a Deep-Learning Framework. Preprint at https://doi.org/10.48550/arXiv.2005.03230 (2020).
80. Millidge, B., Song, Y., Salvatori, T., Lukasiewicz, T. & Bogacz, R. A Theoretical Framework for Inference and Learning in Predictive Coding Networks. in (2022).
81. Tscshantz, A., Millidge, B., Seth, A. K. & Buckley, C. L. Hybrid predictive coding: Inferring, fast and slow. *PLOS Comput. Biol.* **19**, e1011280 (2023).
82. Salvatori, T. *et al.* A Survey on Brain-Inspired Deep Learning via Predictive Coding. Preprint at https://doi.org/10.48550/arXiv.2308.07870 (2025).
83. Friston, K. J., Stephan, K. E., Montague, R. & Dolan, R. J. Computational psychiatry: the brain as a phantastic organ. *Lancet Psychiatry* **1**, 148–158 (2014).
84. Langdon, C., Genkin, M. & Engel, T. A. A unifying perspective on neural manifolds and circuits for cognition. *Nat. Rev. Neurosci.* **24**, 363–377 (2023).
85. Chung, S. & Abbott, L. F. Neural population geometry: An approach for understanding biological and artificial neural networks. *Curr. Opin. Neurobiol.* **70**, 137–144 (2021).
86. Sadtler, P. T. *et al.* Neural constraints on learning. *Nature* **512**, 423–426 (2014).
87. Stringer, C., Pachitariu, M., Steinmetz, N., Carandini, M. & Harris, K. D. High-dimensional geometry of population responses in visual cortex. *Nature* **571**, 361–365 (2019).
88. Hénaff, O. J. *et al.* Primary visual cortex straightens natural video trajectories. *Nat. Commun.* **12**, 5982 (2021).
89. Yoon, I. H. R. *et al.* Tracking the topology of neural manifolds across populations. *Proc. Natl. Acad. Sci.* **121**, e2407997121 (2024).
90. Beiran, M., Meirhaeghe, N., Sohn, H., Jazayeri, M. & Ostojic, S. Parametric control of flexible timing through low-dimensional neural manifolds. *Neuron* **111**, 739-753.e8 (2023).
91. Gallego, J. A. Neural manifolds: more than the sum of their neurons. *Nat. Rev. Neurosci.* **26**, 312–312 (2025).
92. Amsaleg, L. *et al.* The vulnerability of learning to adversarial perturbation increases with intrinsic dimensionality. in *2017 IEEE Workshop on Information Forensics and Security (WIFS)* 1–6 (2017). doi:10.1109/WIFS.2017.8267651.
93. Ma, X. *et al.* Characterizing Adversarial Subspaces Using Local Intrinsic Dimensionality. Preprint at https://doi.org/10.48550/arXiv.1801.02613 (2018).
94. Nassar, J., Sokol, P. A., Chung, S., Harris, K. D. & Park, I. M. On 1/n neural representation and robustness. Preprint at https://doi.org/10.48550/arXiv.2012.04729 (2020).
95. Ansuini, A., Laio, A., Macke, J. H. & Zoccolan, D. Intrinsic dimension of data representations in deep neural networks. Preprint at https://doi.org/10.48550/arXiv.1905.12784 (2019).
96. Alain, G. & Bengio, Y. Understanding intermediate layers using linear classifier probes. Preprint at https://doi.org/10.48550/arXiv.1610.01644 (2018).
97. Cohen, U., Chung, S., Lee, D. D. & Sompolinsky, H. Separability and geometry of object manifolds in deep neural networks. *Nat. Commun.* **11**, 746 (2020).
98. Ehrlich, D. A., Schneider, A. C., Priesemann, V., Wibral, M. & Makkeh, A. A Measure of the Complexity of Neural Representations based on Partial Information Decomposition. Preprint at https://doi.org/10.48550/arXiv.2209.10438 (2023).
99. Dayan, P., Hinton, G. E., Neal, R. M. & Zemel, R. S. The Helmholtz Machine. *Neural Comput.* **7**, 889–904 (1995).
100. Rao, R. P. N., Gklezakos, D. C. & Sathish, V. Active Predictive Coding: A Unifying Neural Model for Active Perception, Compositional Learning, and Hierarchical Planning. *Neural Comput.* **36**, 1–32 (2023).
101. Bastos, A. M. *et al.* Canonical microcircuits for predictive coding. *Neuron* **76**, 695–711 (2012).
102. Zeki, S. & Shipp, S. The functional logic of cortical connections. *Nature* **335**, 311–317 (1988).
103. Zeki, S. The Ferrier Lecture 1995 Behind the Seen: The functional specialization of the brain in space and time. *Philos. Trans. R. Soc. B Biol. Sci.* **360**, 1145–1183 (2005).

104. Shipp, S. Neural Elements for Predictive Coding. *Front. Psychol.* **7**, (2016).
105. Winn, J. & Bishop, C. M. Variational Message Passing. *J. Mach. Learn. Res.* **6**, 661–694 (2005).
106. Penny, W. D. Comparing Dynamic Causal Models using AIC, BIC and Free Energy. *NeuroImage* **59**, 319–330 (2012).
107. Wallace, C. S. & Dowe, D. L. Minimum Message Length and Kolmogorov Complexity. *Comput. J.* **42**, 270–283 (1999).
108. Hinton, G. E. & van Camp, D. Keeping the neural networks simple by minimizing the description length of the weights. in *Proceedings of the sixth annual conference on Computational learning theory* 5–13 (Association for Computing Machinery, New York, NY, USA, 1993). doi:10.1145/168304.168306.
109. Hinton, G. E. & Zemel, R. S. Autoencoders, minimum description length and Helmholtz free energy. in *Proceedings of the 7th International Conference on Neural Information Processing Systems* 3–10 (Morgan Kaufmann Publishers Inc., San Francisco, CA, USA, 1993).
110. Schmidhuber, J. Formal Theory of Creativity, Fun, and Intrinsic Motivation (1990–2010). *IEEE Trans. Auton. Ment. Dev.* **2**, 230–247 (2010).
111. Marino, J. Predictive Coding, Variational Autoencoders, and Biological Connections. *Neural Comput.* **34**, 1–44 (2022).
112. Mescheder, L., Nowozin, S. & Geiger, A. Adversarial Variational Bayes: Unifying Variational Autoencoders and Generative Adversarial Networks. in *Proceedings of the 34th International Conference on Machine Learning* 2391–2400 (PMLR, 2017).
113. Tishby, N., Pereira, F. C. & Bialek, W. The information bottleneck method. in *Proc. of the 37-th Annual Allerton Conference on Communication, Control and Computing* 368–377 (1999).
114. Hafez-Kolahi, H. & Kasaei, S. Information Bottleneck and its Applications in Deep Learning. Preprint at https://doi.org/10.48550/arXiv.1904.03743 (2019).
115. Sengupta, B. & Friston, K. J. How Robust are Deep Neural Networks? Preprint at https://doi.org/10.48550/arXiv.1804.11313 (2018).
116. Rao, R. P. N. A sensory–motor theory of the neocortex. *Nat. Neurosci.* **27**, 1221–1235 (2024).
117. Konkle, T. & Alvarez, G. Cognitive Steering in Deep Neural Networks via Long-Range Modulatory Feedback Connections. *Adv. Neural Inf. Process. Syst.* **36**, 21613–21634 (2023).
118. Kingma, D. P. & Ba, J. Adam: A Method for Stochastic Optimization. in *ICLR* (eds Bengio, Y. & LeCun, Y.) (2015).
119. Krizhevsky, A. & Hinton, G. Learning multiple layers of features from tiny images. (2009).
120. Walther, A. *et al.* Reliability of dissimilarity measures for multi-voxel pattern analysis. *Neuroimage* **137**, 188–200 (2016).